\documentstyle[twoside,fleqn,espcrc2]{article}

\def \bc {\begin{center}}
\def \ec {\end{center}}

\def \bfr {\begin{flushright}}
\def \efr {\end{flushright}}

\def \ba {\begin{array}}
\def \ea {\end{array}}

\def \bea {\begin{eqnarray}}
\def \eea {\end{eqnarray}}

\def \be {\begin{equation}}
\def \ee {\end{equation}}

\def\nn{\nonumber}

\def\f{\frac}
\def\l[{\left[}
\def\r]{\right]}
\def\TG{\tilde{G}}
\def\TT{\tilde{T}}

\def\tg{\tilde{g}}

\def\nlnrs{N_{\lambda\nu\rho\sigma}^{(\pm)}}

\def\kpm{\kappa_{(\pm)}}
\def\kmp{\kappa_{(\mp)}}

\def\dim{{\small {\rm dim}}}

\newcommand{\apm}[1]{a^{\dag(\pm)}_{#1}}
\newcommand{\apmb}[1]{a^{(\pm)}_{#1}}

\newcommand{\fpm}[1]{\varphi^{\dag(\pm)}_{#1}}

\newcommand{\fpmb}[1]{\varphi^{(\pm)}_{#1}}

\newcommand{\A}[1]{A^{(\pm)}_{#1}}

\newcommand{\AmS}{{\protect\the\textfont2
  A\kern-.1667em\lower.5ex\hbox{M}\kern-.125emS}}

\title{Algebraic characterization of constraints and generation of mass in 
gauge theories }

\author{M. Calixto\address{Department of Physics, 
University of Wales Swansea, Singleton Park, 
Swansea, SA2 8PP, U.K. and \\ Instituto Carlos I de F\'\i sica 
Te\'orica y Computacional, Facultad
de Ciencias, Universidad de Granada, Campus de Fuentenueva, 
Granada 18002, Spain.}
        and 
        V. Aldaya\address{Instituto de Astrof\'{\i}sica 
de Andaluc\'{\i}a, Apartado Postal 3004,
18080 Granada, Spain.}%
        \thanks{Work partially
supported by the DGICYT.}}
       
\begin{document}

\begin{abstract}
The possibility of 
non-trivial representations of the gauge group on wavefunctionals 
of a gauge invariant quantum field theory leads 
to a generation of mass for intermediate vector and tensor bosons. The mass 
parameters $m$ show up as central charges in the algebra of constraints, which 
then become of second-class nature. The gauge group coordinates 
acquire dynamics outside the null-mass shell and provide the 
longitudinal field degrees of freedom that massless bosons need to form 
massive bosons. 
\end{abstract}

\maketitle

\section{INTRODUCTION}
In this paper we discuss a new approach to quantum gauge theories, from 
a group-theoretic perspective, in which mass enters the theory 
in a {\it natural} way. More precisely, the presence of mass will manifest 
through non-trivial responses 
\be
U\Psi=D^{(m)}_{\TT}(U)\Psi,\label{gauget}
\ee 
of the wavefunctional $\Psi$ under the action of gauge transformations 
$U\in\TT$, where we denote by $D^{(m)}_{\TT}$ a specific representation of 
the gauge group $\TT$ with index $m$. The standard case $D^{(m)}_{\TT}(U)=1\,, 
\forall U\in \TT$ corresponds to the well-known `Gauss law' condition, which 
also reads $\Phi_{a}\Psi=0$ for infinitesimal gauge transformations 
$U\sim 1+\phi^a \Phi_a$. The case of Abelian representations 
$D^{(\vartheta)}_{\TT}(U_n)=e^{in\vartheta}$ of $\TT$, 
where $n$ denotes the winding 
number of $U_n$, leads to the well-known $\vartheta$-vacuum phenomena. 
We shall see that more general (non-Abelian) representations 
$D^{(m)}_{\TT}$ of the gauge group $\TT$ entail 
{\it non-equivalent quantizations} (in the sense of, e.g. 
\cite{McMullan,Landsman}) and a {\it generation of mass}.
 
This non-trivial response of $\Psi$ under gauge transformations $U$ 
causes a {\it deformation} of the corresponding 
Lie-algebra commutators and leads to 
the appearance of central terms proportional to  mass parameters 
(eventually parametrizing the {\it non-equivalent quantizations}) in
the algebra of constraints, which then become a mixture of first- and 
second-class constraints. As a result,  extra (internal) field degrees 
of freedom emerge out of second-class constraints and are 
transferred to the gauge potentials to conform massive  bosons 
(without Higgs fields!).

Thus, the `classical' case $D^{(m)}_{\TT}=1$ is not in general 
preserved in passing to 
the quantum theory. Upon quantization, first-class constraints (connected 
with a gauge invariance of the classical system) might become second-class, a 
metamorphosis which is familiar when quantizing {\it anomalous} gauge 
theories. Quantum ``anomalies'' change the picture of physical states 
being singlets under the constraint algebra. Anomalous (unexpected) 
situations generally go with the standard viewpoint of quantizing 
classical systems and the avoidance of them is evident 
when quantizing, for example, Yang-Mills theory with chiral fermions, where 
a cancellation of gauge anomalies is apparently needed; however, 
these breakdowns, which sometimes are 
inescapable obstacles for canonical quantization,  could be reinterpreted as 
normal (even essential) situations in a wider setting. 
Dealing with constraints directly in the quantum arena, 
this transmutation in the nature of constraints should be naturally allowed, 
as it provides new richness to the quantum theory.

This {\it cohomological} mechanism 
of mass generation makes perfect sense from a Group Approach to Quantization 
(GAQ \cite{GAQ}) framework, which 
is, at heart, an operator description of a quantum system. Thus, our  
essential ingredient to define a quantum system 
will be a given underlying symmetry algebra $\tilde{{\cal G}}$ 
rather than an action functional $S$, which is the standard starting point 
in the usual (``classically-oriented'') formulation of QFT. 

In order to set the context, let us describe a simple, 
but illustrative, example of an abstract quantizing 
algebra  $\tilde{{\cal G}}$ which eventually applies to a diversity of 
physical systems. 
\section{A SIMPLE ABSTRACT QUANTIZING ALGEBRA}

Our particular algebra under study will be the following: 
\bea
\l[X_j,P_k\r]&=&i\delta_{jk}I\,,\nn\\
\l[\Phi_a,\Phi_b\r]&=&if_{ab}^c\Phi_c+ im_{ab}I\,,\label{thealgebra}\\
\l[X_j,\Phi_a\r]&=&i\check{f}_{ja}^k X_k,\,\,\,
\l[P_j,\Phi_a\r]=i\check{f}_{ja}^k P_k,\nn
\eea
where $X_j$ and $P_k$ represent standard ``position'' and ``momentum'' 
operators, respectively, corresponding to the {\it extended phase space} 
${\cal F}$ of the preconstrained (free-like) theory; 
The operators $\Phi_a$ represent 
the constraints which, for the moment, are supposed to close a Lie subalgebra 
$\tilde{{\cal T}}$ with structure constants $f_{ab}^c$ and central charges 
$m_{ab}$. 
We also consider a diagonal 
action of constraints $\Phi$ on $X$ and $P$ with structure constants 
$\check{f}_{ja}^k$  (non-diagonal actions mixing $X$ and $P$ lead to 
interesting ``anomalous'' situations which we shall not discuss here
\cite{symplin}).  By $I$ we simply denote the identity operator, 
that is, the generator of the typical phase invariance 
$\Psi\sim e^{i\beta}\Psi$ of Quantum Mechanics. At this stage, it is worth
mentioning that we could have introduced dynamics in our model by adding  
a Hamiltonian operator $H$ to $\tilde{{\cal G}}$.  
However, we have preferred not to include it because, although 
we could make compatible the dynamics $H$ and the constraints $\Phi$, the  
price could result in an unpleasant enlarging of $\tilde{{\cal G}}$, 
which would make the quantization procedure much more involved. 
Anyway, for us, the {\it true} dynamics (that which preserves 
the constraints) will eventually arise as part of 
the set of {\it good} operators (observables) of the theory (see below).

Note that a flexibility 
in the class of the constraints has being allowed by introducing arbitrary 
central charges $m_{ab}$ in (\ref{thealgebra}). Thus, 
the operators $\Phi_a$ represent a mixed set of first- and second-class 
constraints. Let us denote by ${\cal T}^{(1)}=\{\Phi^{(1)}_n\}$ 
the subalgebra of first-class constraints, that is, 
the ones which do not give rise 
to central terms proportional to $m_{ab}$ at the right hand side of 
the commutators (\ref{thealgebra}). The rest of constraints (second-class) 
will be arranged by conjugated pairs 
$(\Phi^{(2)}_{+\alpha},\Phi^{(2)}_{-\alpha})$, so that 
$m_{+\alpha,-\alpha}\not=0$. 

The simplest (`classical') case 
is when $m_{ab}=0,\,\,\forall a,b$, that is, when all constraints 
are first class ${\cal T}^{(1)}={\cal T}=\tilde{{\cal T}}/u(1)$ and wave 
functions are singlets under ${\cal T}$. However, the 
`quantum' case $m_{ab}\not=0$ entails non-equivalent quantizations 
with important physical consequences. This possibility indicates a 
non-trivial response (\ref{gauget}) of the wave function $\Psi$ under 
$\tilde{{\cal T}}$. That is, $\Psi$ 
acquires a non-trivial dependence on extra degrees of freedom 
$\phi^{(2)}_{-\alpha}$ (`negative modes' attached to pairs of 
second-class constraints), 
in addition to the usual configuration space 
variables $x_j$ (attached to $X_j$). 

Let us formally outline the actual construction 
of the unitary irreducible representations of the group $\TG$ 
with Lie-algebra (\ref{thealgebra}). Wave functions $\Psi$ are defined as 
complex functions on $\TG$, $\Psi:\TG\rightarrow C$, so that the 
(let us say) left-action 
\be
L_{\tg'}\Psi(\tg)\equiv\Psi(\tg' *\tg),\,\,\tg',\tg\in\TG \label{repre}
\ee
defines a reducible (in general) representation of $\TG$. 
The reduction is achieved by means of that maximal set of  
right restrictions on wave functions 
\be
R_{\tg_p}\Psi=\Psi,\,\, \forall 
\tg_p\in G_p,\,\label{pola}
\ee
(which commute with the left action)  
compatible with the natural condition $I\Psi=\Psi$.  
The right restrictions (\ref{pola}) 
generalize the notion of {\it polarization conditions} 
of Geometric Quantization and give rise to a certain representation space 
depending on the choice of the subgroup $G_p\subset\TG$. For the algebra 
(\ref{thealgebra}), a polarization subgroup can be 
$G_p^{(P)}=F_P\times_sT_p$, that is, the semi-direct product of 
the Abelian group of translations $F_P$ generated by 
${\cal F}_P\equiv\{P_k\}$ (half of the symplectic generators in ${\cal F}$) 
by a polarization subalgebra ${\cal T}_p=\{\Phi^{(1)}_n,
\Phi^{(2)}_{+\alpha}\}$ 
of $\TT$ consisting of first-class constraints and half of 
second-class constraints (namely, the `positive modes'). The polarization 
conditions (\ref{pola}) lead to the configuration-space representation 
made of wave functions $\Psi(x_j,\phi^{(2)}_{-\alpha})$ 
depending arbitrarily on 
the group coordinates on $\TG/G_p$ only. Thus, 
as mentioned above, wave functions transform non-trivially  
under the left-action $L_\phi\Psi(\tg)=
D_{\TT}^{(m)}(\phi)\Psi(\tg)$ of $\TT$ according to a given 
representation $D_{\TT}^{(m)}$ 
like in (\ref{gauget}). The physical Hilbert space is made of those wave 
functions $\Psi_{{\small{\rm ph.}}}$ 
that transform as `highest-weight vectors' under $\tilde{T}$, that is, 
they stay invariant under the left-action of first-class constraints and 
(let us say) negative second-class modes: 
\bea
L_{\phi^{(1)}_n}\Psi_{{\small{\rm ph.}}}\!\!\!\!&=&\!\!\!\! 
\Psi_{{\small{\rm ph.}}}\,,\; n=1,..., 
\dim(T^{(1)}),\nn\\ 
L_{\phi^{(2)}_{-\alpha}}\Psi_{{\small{\rm ph.}}}\!\!\!\!&=&\!\!\!\!
\Psi_{{\small{\rm ph.}}}
\,, \; 
\alpha=1,...,\dim(T/T^{(1)})/2\,,
\label{tpcons}
\eea
which close the subgroup $T_p\subset\TT$.

The counting of {\it true 
degrees of freedom} is as follows: polarized-constrained 
wave functions (\ref{tpcons}) depend 
arbitrarily on $d=\dim(\TG)-\dim(G_p)-\dim(T_p)-1$ reduced-space 
coordinates (we are subtracting the phase coordinate $e^{i\beta}$). 
The algebra 
of observables of the theory, $\tilde{{\cal G}}_{\small{\rm good}}
\subset{\cal U}(\tilde{{\cal G}})$ (the enveloping algebra),  
has to be found inside the {\it normalizer} of constraints, that is: 
\be
\l[\tilde{{\cal G}}_{\small{\rm good}},{\cal T}_p\r]\subset {\cal T}_p\,.
\ee
From this characterization, the subalgebra of first-class constraints 
${\cal T}^{(1)}$ become a horizontal ideal (a {\it gauge} subalgebra 
\cite{config}) of $\tilde{{\cal G}}_{\small{\rm good}}$. The Hamiltonian 
operator has to be found inside $\tilde{{\cal G}}_{\small{\rm good}}$ 
by using extra physical arguments.

In what follows, the quantization of massless and 
massive non-Abelian Yang-Mills, 
linear Gravity and Abelian two-form gauge field theories are developed 
from this new approach, where a cohomological origin of mass is pointed out.

\section{UNIFIED QUANTIZATION OF MASSLESS AND MASSIVE VECTOR AND 
TENSOR BOSONS}
  
Let us start with the simplest case of the electromagnetic field. Let us use 
a Fourier parametrization
\bea
A_{\mu}(x)&\equiv & \int  \f{d^3k}{2k^0}[a_{\mu}(k)e^{-ikx}+
a^\dag_{\mu}(k)e^{ikx}]\,,\nn\\
\Phi(x)&\equiv & \int  \f{d^3k}{2k^0}[ \varphi(k)e^{-ikx}+
 \varphi^\dag(k)e^{ikx}]\,,\label{fourierpar}
\eea
for the vector potential $A_\mu(x)$ and the constraints $\Phi(x)$ 
(the generators of local $U(1)(x)$ gauge transformations). The Lie algebra 
$\tilde{{\cal G}}$ of the quantizing electromagnetic group $\TG$ has the 
following form \cite{empro}
\bea
\left[a_{\mu}(k),a^\dag_{\nu}(k')\right]&=&
 \eta_{\mu\nu}\Delta_{kk'}I\,, \nn\\
 \left[\varphi(k),
\varphi^\dag(k')\right]&=&k^2\Delta_{kk'}I\,, \nn\\ 
\left[a^\dag_{\mu}(k),\varphi(k')\right]&=&
-ik_{\mu}\Delta_{kk'}I \,,\nn\\
\left[a_{\mu}(k),
\varphi^\dag(k')\right]&=&-ik_{\mu}\Delta_{kk'}I \,,
\label{conmutadores}
\eea
where $\Delta_{kk'}= 2k^0 \delta^3(k-k')$ 
is the generalized delta function on the positive 
sheet of the mass hyperboloid and $k^2=m^2$ is the squared mass. Constraints 
are first-class for $k^2=0$ and constraint equations 
$\varphi\Psi=0=\varphi^\dag\Psi$ keep 2 field degrees of freedom out of 
the original 4, as corresponds to a photon. For $k^2\not=0$, constraints are 
second-class and the restrictions $\varphi\Psi=0$ keep 3 field degrees 
of freedom out of the original 4, as corresponds to a Proca field.

For symmetric and anti-symmetric tensor potentials $\A{\mu\nu}$, the 
algebra is the following \cite{gtp}:
\bea
\left[\apmb{\lambda\nu}(k),
\apm{\rho\sigma}(k')\right]&=&
 \nlnrs\Delta_{kk'}I\,, \nn \\ 
\left[\fpmb{\rho}(k),
\fpm{\sigma}(k')\right]&=&k^2M^{(\pm)}_{\rho\sigma}(k)
\Delta_{kk'}I\,,\nn\\ 
\left[\apm{\lambda\nu}(k),\fpmb{\sigma}(k')\right]&=&
-ik^{\rho}\nlnrs\Delta_{kk'}I\,,\nn\\
 \left[\apmb{\lambda\nu}(k),
\fpm{\sigma}(k')\right]&=&-ik^{\rho}\nlnrs\Delta_{kk'}I\,,
\label{conmutadores1}
\eea
where $M^{(\pm)}_{\rho\sigma}(k) \equiv \eta_{\rho\sigma} -
\kmp\f{k_\rho k_\sigma}{k^2}$ and   
$\nlnrs \equiv \eta_{\lambda\rho}\eta_{\nu\sigma} \pm   
\eta_{\lambda\sigma}\eta_{\nu\rho} - 
\kpm\eta_{\lambda\nu}\eta_{\rho\sigma}$,   
with $\kappa_{(+)}=1$ and $\kappa_{(-)}=0$. For the massless  
$k^2=0$ case, all constraints are first-class for the symmetric case, 
whereas the massless, anti-symmetric case 
possesses  a couple of second-class constraints:
\be
\left[\check{k}^\rho\varphi^{(-)}_{\rho}(k),
{\check{k}'}{}^\sigma\varphi^{(-)\dag}_{\sigma}(k')\right]=4(k^0)^4\Delta_{kk'}
I\,,\label{2ndclass}
\ee
where $\check{k}^\rho\equiv k_\rho$. Thus, first-class constraints for the 
massless anti-symmetric case are ${\cal T}_{(-)}^{(1)}=
\{\epsilon_\mu^\rho\varphi^{(-)}_{\rho},\,
\epsilon_\mu^\rho\varphi^{(-)\dag}_{\rho}\},\,
\mu=0,1,2,\,$ where $\epsilon_\mu^\rho$ is a tetrad which diagonalizes the 
matrix $P_{\rho\sigma}=k_\rho k_\sigma$; in particular, we choose 
$\epsilon_3^\rho\equiv \check{k}^\rho$ and 
$\epsilon_0^\rho\equiv k^\rho$. There are $2=10-8$ true degrees 
of freedom for the symmetric case (a massless graviton) and $1=6-5$ for the 
anti-symmetric case (a pseudo-scalar particle).

For $k^2\not=0$, all constraints are second-class for the 
symmetric case,  whereas, for the 
anti-symmetric case,  constraints close a Proca-like subalgebra which 
leads to three pairs of second-class 
constraints, and a pair  
of gauge vector fields $(k^\lambda\varphi^{(-)}_{\lambda},\,
k^\lambda\varphi^{(-)\dag}_{\lambda})$. The constraint equations keep 
$6=10-4$ field degrees of freedom for the symmetric case 
(massive spin 2 particle 
+ massive scalar field ---the trace of the symmetric tensor), 
and $3=6-3$ field degrees of freedom for 
the anti-symmetric case (massive pseudo-vector particle).

For non-Abelian $SU(N)$ Yang-Mills theories in the Weyl gauge $A^0=0$ 
there is still a residual gauge invariance $T={\rm Map}(\Re^3,SU(N))$. 
The basic commutators between the non-Abelian vector potentials 
$A^j_a(x),\,j=1,2,3; a=1,...,N^2-1$, the electric field $E^j_a(x)$ and 
the (Gauss law) constraints $\Phi_a(x)$ are \cite{ymas}
 \bea
\l[{A}^j_a(x), {E}^k_b(y)\r]&=&
i\delta_{ab}\delta^{jk}\delta(x-y){I}\,,\nn\\
\l[{\Phi}_a(x), {\Phi}_b(y)\r]&=&-if_{ab}^c\delta(x-y) 
{\Phi}_c(x)\nn\\ 
& &-if_{ab}^c\frac{\lambda_c}{r^2}\delta(x-y){I}\,,
\nn\\
\l[{A}^j_a(x), {\Phi}_b(y)\r]&=&
-if_{ab}^c\delta(x-y) {A}^j_c(x)\nn\\
& & -\frac{i}{r}\delta_{ab}\partial^j_x\delta(x-y){I}
\,,\nn\\
\l[{E}^j_a(x), {\Phi}_b(y)\r]&=&-if_{ab}^c\delta(x-y) 
{E}^j_c(x)\,,\label{YM}
\eea
where $r$ is the coupling constant and $\lambda_{ab}=f^c_{ab}\lambda_c$ is 
a mass matrix $(\lambda\sim m^3$). Let us denote by 
$c\equiv{\rm dim}(T^{(1)})$ and $\tau\equiv N^2-1$ the dimensions of the 
rigid subgroup of first-class constraints and $SU(N)$ respectively. 
Unpolarized wave functions 
$\Psi(A^j_a,E^j_a,\phi_a)$ depend on $n=2\times 3\tau+\tau$ field 
coordinates in $d=3$ dimensions; polarization equations introduce $p=c+
\frac{n-c}{2}$ independent restrictions on 
wave functions, corresponding to $c$ non-dynamical coordinates in $T^{(1)}$ 
and half of the dynamical ones; finally, constraints (\ref{tpcons}) impose 
$q=c+\frac{\tau-c}{2}$ additional restrictions which leave 
$f=n-p-q=2c+3(\tau-c)$ field degrees of freedom (in $d=3$). 
 These fields correspond to 
$c$ massless vector bosons (2 polarizations) attached to ${T}^{(1)}$ and 
$\tau-c$ massive vector bosons.  In particular, for the massless 
case, we have $c=\tau$,  
since constraints are {\it first-class} 
(that is, we can impose $q=\tau$ restrictions) and constrained wave 
functions have support on $f_{m=0}=3\tau-\tau=2\tau\leq f_{m\not=0}$ 
arbitrary fields corresponding to $\tau$ massless vector bosons. 
The subalgebra ${\cal T}^{(1)}$ corresponds to the unbroken 
gauge symmetry of the constrained theory. There are distinct 
symmetry-breaking patterns $\tilde{\cal T}\rightarrow {\cal T}^{(1)}$ 
according to the different choices of mass-matrices 
$\lambda_{ab}=f^c_{ab}\lambda_c$ in (\ref{YM}).

\end{document}